# Frequency-resolved ultrafast electron diffraction: visualizing vibrational dynamics in frequency- and real-space

Rosalie Tabarie[a], Simon P. Neville[b], Michael Schuurman[b,c], and Kasra Amini[a,*]

[a]Max-Born-Institut, Max-Born-Str. 2A, 12489, Berlin, Germany.

[b]National Research Council Canada 100 Sussex Drive Ottawa, Canada, K1A 0R6.

[c]Department of Chemistry and Biomolecular Sciences University of Ottawa 10 Marie Curie Pvt Ottawa, Canada, K1N 6N5.

*Corresponding author email: kasra.amini@mbi-berlin.de

## Abstract

Ultrafast electron diffraction (UED) provides direct information on changes in molecular structure following photoexcitation. However, identifying the individual vibrational motions contributing to these structural dynamics remains challenging from time-dependent electron scattering and pair distribution functions alone, particularly when multiple vibrational motions contribute over similar internuclear distances. Moreover, the ~100 fs temporal resolution of current gas-phase UED experiments further limits the vibrational dynamics that can be resolved. In this work, we introduce frequency-resolved UED, where Fourier transformation of the time-dependent difference pair distribution function ($\Delta\mathrm{PDF}$) along the pump-probe delay axis gives a two-dimensional frequency-distance representation of the photoinduced structural dynamics. To demonstrate the utility of frequency-resolved UED, we study the ultrafast vibrational dynamics in allene and its methylated derivative 1,2-butadiene following photoexcitation to its $S_1(\pi\pi^*)$ state at 200 nm. We simulate the electron scattering signals from previously-published *ab initio* multiple spawning (AIMS) trajectories (S. P. Neville *et al.*, *J. Chem. Phys.*, 2016, **144**, 014305). Through the frequency-distance representation, we identify the C=C stretching and CCC bending motions in both molecules and their internuclear distances over which these frequency components contribute. We further separate overlapping $CH_2$ vibrational motions and identify contributions from multiple vibrational frequencies at the same internuclear distance. We show the delay times at which specific frequency components contribute during the excited-state dynamics by changing the pump-probe delay range used for the Fourier transform of the $\Delta\mathrm{PDF}$ signal. In both molecules, the CCC bending motion contributes predominantly before substantial $S_1 \rightarrow S_0$ population transfer occurs by 70-90 fs, whereas the C=C stretching motion persists at all delays. We find that methyl substitution reduces the C=C stretching and CCC bending frequencies, while an additional frequency component appears in 1,2-butadiene and contributes primarily during the first 90 fs. Finally, we investigate the impact of the total instrument response function (IRF) on the retrieval of these frequency components and show the importance of reaching sub-20-fs, and ultimately few-femtosecond, temporal resolution. Frequency-resolved UED therefore provides a route to identify the vibrational frequencies that contribute to photoinduced structural dynamics, the internuclear distances associated with these frequencies, and the reaction times at which they are present.

## Introduction

The outcome of many photochemical reactions can be strongly influenced by the nuclear wave packet (NWP) dynamics that can occur as fast as the few-femtosecond1–3 to few-tens-of-femtosecond timescale.4–9 This is particularly important near conical intersections (CIs), where the NWP can evolve along specific nuclear coordinates that impact both the electronic population dynamics and how the reaction branches into different products.10–13 Directly measuring these coupled electronic-nuclear dynamics remains a major challenge in ultrafast photochemistry, since some of the fastest nuclear motions occur on timescales shorter than the instrument response function (IRF) of many ultrafast experiments and can therefore be averaged out. To understand its excited-state molecular dynamics, and thus the underlying reaction mechanism, it is crucial to identify the vibrational motions involved as the NWP evolves towards and through a CI.

Visualizing changes in molecular structure and excited-state dynamics on their natural vibrational time (i.e., femtosecond) and spatial (i.e., Ångström) scales has been investigated using a range of ultrafast spectroscopic and imaging techniques. Ultrafast spectroscopy[5,9,12,14–16] provides powerful methods for measuring the electronic and vibrational dynamics that occur following photoexcitation. In particular, multidimensional spectroscopies[17] such as two-dimensional infrared (2D-IR) spectroscopy[18–24] can separate spectral contributions along two frequency axes, allowing correlations between different vibrational modes to be identified. For example, cross-peak signals in 2D-IR can provide information on vibrational coupling and valuable structural information, since these couplings depend on the relative atomic positions and orientations of specific groups or parts of the molecular structure.[19,25] Other multidimensional spectroscopies are particularly suited to studying photoinduced excited-state and coupled electronic-nuclear dynamics. For excited-state dynamics, two-dimensional electronic spectroscopy (2DES) can resolve electronic couplings, coherences, and population dynamics following photoexcitation.[26,27] While correlations between electronic excitation and the vibrational response of the photoexcited molecule can be probed through two-dimensional electronic-vibrational (2DEV) spectroscopy, combining electronic excitation with vibrational detection.[28–30]

While structural information is obtained from the measured vibrational frequencies and couplings, 2D-IR does not directly provide an observable of signal changes occurring at specific internuclear distances. Structural imaging techniques[31–47] are complementary to spectroscopic methods as they measure changes in molecular structure after photoexcitation. For example, in ultrafast electron diffraction (UED),[32–42] the time-dependent molecular interference signal can be sine transformed into a pair distribution function (PDF),[48] allowing changes in molecular structure to be visualized as a function of internuclear distance, $R$, and pump-probe delay, $t$. The difference in the pair distribution function, $\Delta\mathrm{PDF}(R,t)$, therefore shows the internuclear distance regions over which structural changes occur during the excited-state dynamics. However, conventional UED measurements are generally analysed in time and real-space coordinates, and do not readily identify the vibrational frequencies that contribute to the observed signal changes at each internuclear distance.

Here, in this work, we introduce frequency-resolved UED, where the time-dependent oscillations in the $\Delta\mathrm{PDF}(R,t)$ signal are Fourier transformed along the pump-probe delay to obtain a two-dimensional frequency-distance map. In this frequency-distance representation, the vibrational frequencies that contribute to the structural dynamics can be resolved and, importantly, the internuclear distance regions over which they appear. This provides the possibility to correlate the measured vibrational frequencies with real-space changes in molecular structure during a photochemical reaction. Moreover, we also show that we can isolate the reaction time ranges at which the frequency components are present in the frequency-distance map by changing the pump-probe delay time window used for the Fourier transform. We demonstrate this frequency-resolved UED approach using simulated UED signals calculated from previously published *ab initio* multiple spawning (AIMS) trajectories of allene and its methylated derivative, 1,2-butadiene (1,2-BD), following excitation to the $S_1$ state (see Neville *et al.*[49]). In the remainder of this section, we first provide a brief historical overview and current state of gas-phase UED together with an introduction to the photochemistry of allenes (section A), followed by a discussion of the factors that currently limit the IRF in UED (section B), and finally an introduction to the elastic electron scattering theory used in this work (section C).

### A. Ultrafast electron diffraction imaging of gas-phase molecular dynamics

Ultrafast gas-phase electron diffraction was pioneered by Zewail and co-workers in the 1990s, where picosecond electron pulses were used to capture transient molecular structures during photochemical reactions.[36–38] Significant efforts in the UED field have led to the development of shorter electron pulses and have moved gas-phase UED into the femtosecond regime.[33,39–42,50–56] Both keV and MeV UED have been utilized to visualize ultrafast structural dynamics in a range of photochemical reactions, including bond dissociation,[39,40,55,56] ring-opening,[40,42,57] and structural dynamics associated with nonadiabatic relaxation through CIs[39,41]. These measurements have demonstrated the capability of UED to track large changes in molecular structure in real time. However, resolving the much faster individual bending and stretching vibrations remains challenging.

For example, photoexcitation of allene at 200 nm to its bright $S_1(\pi\pi^*)$ state leads to the NWP leaving the Franck-Condon region and twisting of the terminal $CH_2$ groups around the C=C=C axis (Fig. 1a).[49] Specific one-dimensional cuts through the $S_0$ and $S_1$ potential energy surfaces (PESs) of allene along the terminal-group twisting angle, $\psi$, defined by the relative dihedral angle of the two terminal $CH_2$ groups, and the CCC bending coordinate, $\chi$, are shown in Figs. 1b and 1c for the ground-state, $S_0$, (blue) and the first excited state, $S_1$ (red). It is important to note that the $S_1$ PES does not possess local minima, and therefore harmonic frequencies cannot be calculated for allene in its $S_1$ state. Nevertheless, the one-dimensional cuts provide potential energy curves (PECs) along the individual nuclear coordinates and show minimum-energy points along these specific coordinates.

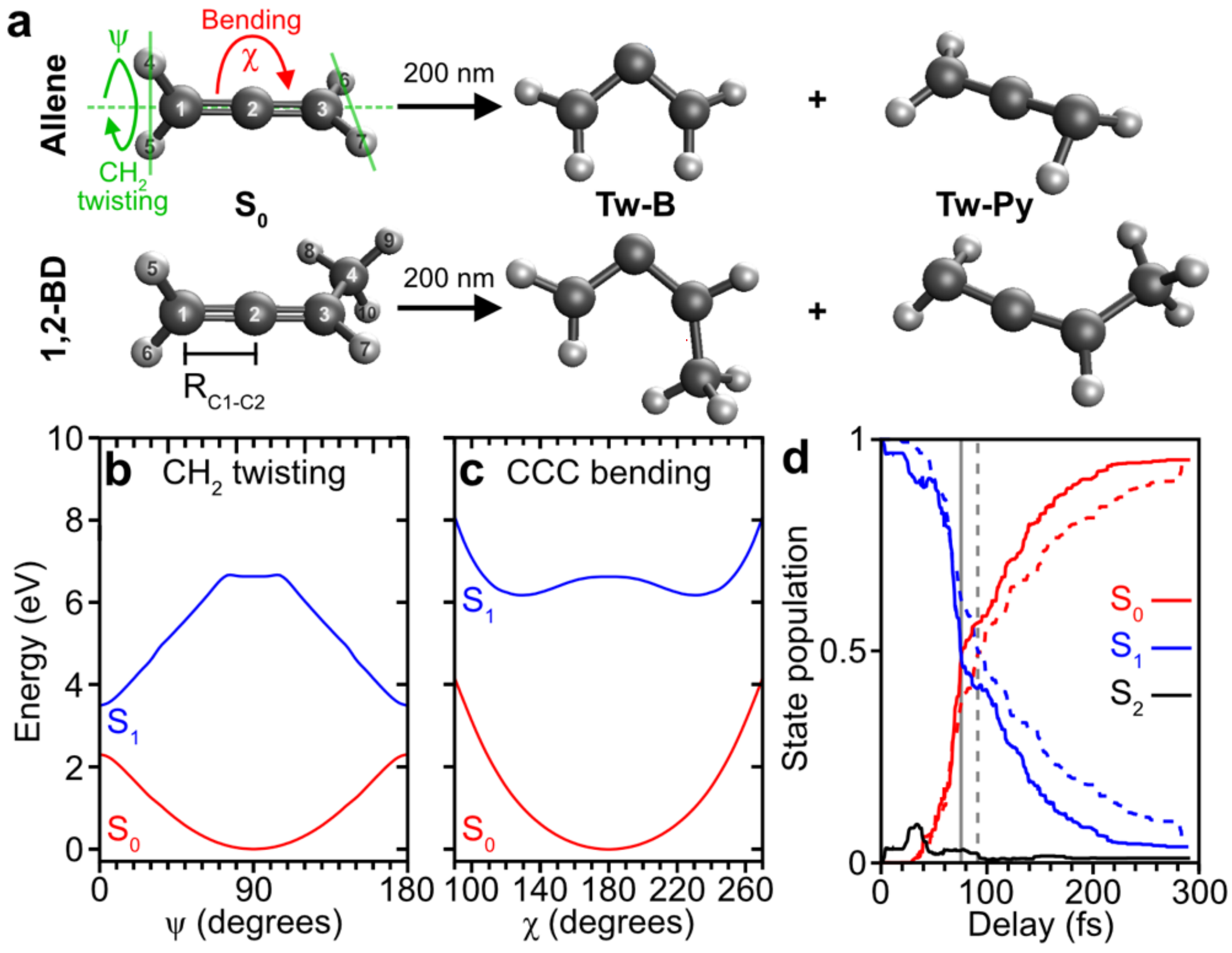

**Fig. 1** **(a)** Optimized geometric structures of allene (top) and 1,2-butadiene (1,2-BD; bottom) in their $S_0$ ground-state and $S_1(\pi\pi^*)$ first excited state with twisting-bending (Tw-B) and twisting-pyramidalization (Tw-Py) coordinates. The angles describing CCC bending, $\chi$, and $CH_2$ twisting, $\psi$, are indicated together with the C1-C2 internuclear distance, $R_{\mathrm{C1-C2}}$. **(b-c)** Potential energy curves (PECs) of allene as a function of the (b) $CH_2$ twisting angle, $\psi$, and (c) CCC bending angle, $\chi$, for the ground state, $S_0$ (red), and the first excited state, $S_1$ (blue). These curves are one-dimensional cuts through the multidimensional potential energy surfaces. Each cut varies only $\psi$ or $\chi$, with all remaining nuclear coordinates fixed at their Franck-Condon values ($\chi = 180°, \psi = 90°$). **(d)** Electronic state population dynamics of the ground state, $S_0$ (red), and the first two excited states, $S_1$ (blue) and $S_2$ (black), for allene (solid) and 1,2-butadiene (dashed). Vertical lines indicate the time required for the $S_1$ population to decrease to 50% for allene (solid) and 1,2-butadiene (dashed). The optimized structures, electronic state populations and PECs were adapted from Ref. [49], with the permission of AIP Publishing.

Along the $CH_2$ twisting coordinate, the $S_1$ PEC reaches minimum-energy points at $\psi = 0°$ and $180°$ (Fig. 1b). Similarly, the $S_1$ PEC along the bending coordinate, $\chi$, at the Franck-Condon geometry also reaches minimum-energy points near $\chi \approx 128°$ and $218°$ (Fig. 1c), displaced from the linear $\chi = 180°$ ground-state geometry. At these $\psi$ angles, the two terminal $CH_2$ groups are coplanar to one another, significantly deviating from their perpendicular orientation at $\psi = 90°$ in the $S_0$ ground state. This twisting motion brings the molecule towards regions of the $S_1/S_0$ CI seam, where internal conversion can proceed through two pathways of either terminal-carbon pyramidalization (Tw-Py) or bending of the CCC backbone (Tw-B) back to the $S_0$ state.[49] The optimized structures of the Tw-Py and Tw-B $S_1/S_0$ at the minimum energy conical intersections (MECIs) are shown in Fig. 1a. Moreover, the $S_1$ population in allene decreases to 50% after approximately 70 fs (vertical grey solid line, Fig. 1d). During this time, key vibrational motions, including $CH_2$ wagging to CCC bending, have vibrational periods of approximately 20 – 60 fs.[49] Methylated derivatives of allene, for example, 1,2-butadiene (1,2-BD), can exhibit inertial effects, with substitution of a terminal hydrogen atom by a methyl group slowing the nuclear dynamics.[49] This methyl substitution increases the time for the $S_1$ population to decrease to 50% in 1,2-BD (~90 fs; vertical grey dashed line, Fig. 1d). Therefore, allene and its methylated derivative 1,2-BD provide useful benchmark systems for examining how UED can resolve the fast vibrational dynamics associated with the competing relaxation pathways. In particular, we investigate their frequency-distance signatures and the IRF required to resolve these signals.

### B. Towards few-fs ultrafast electron diffraction: the IRF challenge

The temporal resolution of gas-phase UED has improved significantly over the past decade, with recent measurements reaching approximately 80 fs using compressed MeV electron beams.[42] This compares with 150-300 fs for uncompressed MeV beams[52] and 240 fs (FWHM) for compressed keV beams[53]. All temporal resolutions mentioned in this article are full width at half maximum (FWHM) values, unless otherwise stated.

To visualize nuclear dynamics that occur on the tens-of-femtosecond timescale or faster, the temporal resolution of UED must therefore be improved further towards the sub-20 fs regime, and ideally into the few-femtosecond regime for resolving the fastest vibrational motions. The temporal resolution of a pump-probe UED experiment is given by the total instrument response function (IRF), $\sigma_{\mathrm{tot}}$, and can be expressed as[58,59]

$$\sigma_{\mathrm{tot}} = \sqrt{\sigma_{\mathrm{e^-}}^2 + \sigma_{\mathrm{pump}}^2 + \sigma_{\mathrm{VM}}^2 + \sigma_{\mathrm{jitter}}^2}, \qquad (1)$$

where $\sigma_{\mathrm{e^-}}$ is the electron pulse duration, in which 25 – 90 keV electrons can be compressed to 50 – 150 fs (FWHM) with RF[60–63] and THz[64–66] compression fields, while 3 MeV beams can be compressed to as low as 8 fs and 29 fs (FWHM) using static magnetic bunch compression with an $\alpha$-magnet[67] and double-bend achromat (DBA)[68], respectively. The pump pulse duration is given by $\sigma_{\mathrm{pump}}$ and often has values of 30-100 fs. The velocity mismatch, $\sigma_{\mathrm{VM}}$, between a weakly-relativistic keV electron probe and optical pump pulses can broaden the temporal response by up to ~1 ps for a 200 – 300 μm gas jet probed by 90 keV electrons. This velocity mismatch can be corrected by tilting the pump pulse front to match the electron group velocity across the gas jet length, or substantially reduced by using highly-relativistic MeV electrons[52,67,68]. Finally, $\sigma_{\mathrm{jitter}}$ is the total timing jitter between the pump and probe pulses, and contains several contributions, as defined by

$$\sigma_{\mathrm{jitter}} = \sqrt{\sigma_{\mathrm{optical}}^2 + \Delta t_{\mathrm{short-term}}^2 + \Delta t_{\mathrm{long-term}}^2 + \Delta t_{\mathrm{HV}}^2}, \qquad (2)$$

where $\sigma_{\mathrm{optical}}$ is the all-optical timing jitter, which is typically on the order of 10-20 fs or better, while $\Delta t_{\mathrm{short-term}}$ and $\Delta t_{\mathrm{long-term}}$ represent the short-term and long-term drifts in time-zero, $t_0$. For RF compression, fluctuations in the phase and amplitude of the compression field can contribute to these timing variations,[61,62,69] with short-term timing jitter corrected to approximately 6 fs and long-term timing drifts of 60-90 fs demonstrated by Amini *et al.*[63]. In comparison, THz compression is inherently synchronized to the driving laser, with a substantially smaller timing jitter associated with the compression field.[70] Finally, $\Delta t_{\mathrm{HV}}$ represents changes in the electron time-of-arrival caused by voltage fluctuations in the accelerator high voltage (HV) power supply, which can correspond to approximately 15 fs (FWHM) for a 0.5 V (FWHM) jitter.

Achieving a sub-20 fs IRF in UED is becoming increasingly feasible. Firstly, μJ-level ultraviolet pulses with few-fs durations can now be generated through resonant dispersive wave (RDW) generation in gas-filled hollow-core fibers[71–73], following seminal studies by Nagy and coworkers[74,75]. Secondly, THz electron compression of keV electron beams to trains of sub-fs pulses has been demonstrated experimentally[76], together with the high timing stability enabled by the optical synchronization of the THz compression field. Moreover, as mentioned earlier, static magnetic compression of a collimated 3 MeV beams to 8 fs was also demonstrated, with a timing jitter of 10 fs.[67] In addition, a 3 MeV has a velocity mismatch of only 3 fs across a 100 μm gas jet. Having considered the temporal resolution required to resolve these fast vibrational motions, we next describe how changes in molecular structure are encoded in the elastic electron scattering signal.

## C. Scattering theory

In an ultrafast electron diffraction experiment, an electron "probe" pulse with initial momentum $\boldsymbol{k_0}$ scatters from a molecule containing $N$ atoms that has been photoexcited by a laser "pump" pulse. The electron momentum after scattering is given by $\boldsymbol{k}$, with $|\boldsymbol{k}| = |\boldsymbol{k_0}|$ under elastic scattering conditions. The magnitude of the momentum transfer, $s$, is given by[35,48]

$$s = |\boldsymbol{k_0} - \boldsymbol{k}| = 2|\boldsymbol{k_0}|\sin\left(\frac{\theta}{2}\right) = \left(\frac{4\pi}{\lambda}\right)\sin\left(\frac{\theta}{2}\right), \qquad (3)$$

where θ is the scattering angle, and λ is the de Broglie wavelength of the probe electrons. The elastic scattering from each atom $i$ is described by a complex elastic scattering amplitude,[35,48]

$$f_{\mathrm{i}}(s) = |f_{\mathrm{i}}(s)|\exp[i\eta_{\mathrm{i}}(s)], \qquad (4)$$

where $|f_{\mathrm{i}}(s)|$ is the magnitude of the elastic scattering amplitude and $\eta_{\mathrm{i}}(s)$ is the corresponding scattering phase. Within the independent atom model (IAM), a molecule is approximated as a collection of isolated atoms, such that the total elastic scattering intensity, $I_{\mathrm{T}}(s)$, can be separated into three contributions. The first contribution arises from scattering from individual atoms and therefore contains no information on the molecular geometry. This contribution is referred to as the incoherent atomic scattering intensity, $I_{\mathrm{A}}(s)$, and is given by[35,48]

$$I_{\mathrm{A}} = \sum_{\mathrm{i}=1}^{\mathrm{N}}|f_{\mathrm{i}}(s)|^2, \qquad (5)$$

where $N$ is the number of atoms in the molecule. The second contribution arises from interference between the elastically scattered waves from pairs of atoms. The incoming electron plane wave generates a spherical wave at each atom, leading to constructive and destructive interference between the waves scattered from each atomic pair. For an isotropic ensemble of molecules, orientational averaging gives the coherent molecular scattering signal, $I_{\mathrm{M}}$(s),[35,48] given by

$$I_{\mathrm{M}}(s) = \sum_{\substack{\mathrm{i,j}=1 \\ \mathrm{i}\neq\mathrm{j}}}^{\mathrm{N}}|f_{\mathrm{i}}(s)||f_{\mathrm{j}}(s)|\cos\left(\eta_{\mathrm{i}}(s) - \eta_{\mathrm{j}}(s)\right)\frac{\sin(s\cdot R_{\mathrm{ij}})}{s\cdot R_{\mathrm{ij}}}, \qquad (6)$$

where $R_{\mathrm{ij}}$ is the internuclear distance between atoms $i$ and $j$. The third contribution arises from additional background contributions, $I_{\mathrm{Bkg}}(s)$, related to detector sensitivity issues, fluctuations in the pump laser pulse, and instabilities in the gas jet source. The total elastic scattering intensity measured in a diffraction experiment is therefore

$$I_{\mathrm{T}}(s,t) = I_{\mathrm{A}}(s) + I_{\mathrm{M}}(s,t) + I_{\mathrm{Bkg}}(s). \quad (7)$$

The time dependence of $I_{\mathrm{M}}(s,t)$ arises from changes in the internuclear distances, $R_{\mathrm{ij}}(t)$, during the nuclear dynamics of a photochemical reaction. These contributions decrease significantly with increasing momentum transfer because the elastic scattering cross-section decreases strongly with increasing scattering angle.[77] The molecular scattering intensity additionally contains the $\sin(s \cdot R_{\mathrm{ij}})/s \cdot R_{\mathrm{ij}}$ interference term, which further reduces its amplitude towards higher momentum transfer.

The molecular structure information is contained in the molecular scattering intensity and is typically extracted using the molecular contrast factor (MCF),

$$M(s) = \frac{I_{\mathrm{M}}(s)}{I_{\mathrm{A}}(s)}. \quad (8)$$

The MCF is subsequently multiplied by $s$, giving the modified molecular scattering signal, $sM$(s), which compensates for the decreasing amplitude of the molecular interference oscillations towards larger momentum transfer. In pump-probe UED experiments, the structural dynamics are measured through the time-dependent relative difference signal referenced to the unpumped molecule,[41]

$$\Delta I/I_0(s,t) = \frac{I_{\mathrm{T}}(s,t) - I_{\mathrm{T}}(s,t<0)}{I_{\mathrm{T}}(s,t<0)}, \quad (9)$$

from which the corresponding changes in the modified molecular scattering signal can be obtained. The modified molecular scattering difference signal is defined as[40]

$$\Delta sM(s,t) = s[M(s,t) - M(s,t<0)]. \quad (10)$$

A sine transform of $\Delta sM(s,t)$ over the experimentally accessible momentum transfer range, between $s_{\mathrm{min}}$ and $s_{\mathrm{max}}$, yields the time-dependent difference pair distribution function,[40]

$$\Delta \mathrm{PDF}(R,t) = \int_{s_{\mathrm{min}}}^{s_{\mathrm{max}}} \Delta sM(s,t) \sin(sR) \exp(-\alpha s^2)\, \mathrm{d}s, \quad (11)$$

which maps the structural changes into real space and provides information on changes in internuclear distances during the reaction. The Gaussian damping term, $\exp(-\alpha s^2)$, suppresses Fourier truncation artefacts arising from the finite momentum transfer range. Consequently, $\Delta\mathrm{PDF}(R,t)$ provides a visualization of the structural evolution following photoexcitation. Oscillatory features in both $\Delta I/I_0(s,t)$ and $\Delta\mathrm{PDF}(R,t)$ can additionally contain information on coherent vibrational motion. Fourier transformation of the $\Delta\mathrm{PDF}(R,t)$ signal along the pump-probe delay axis yields a two-dimensional frequency-distance map, showing the internuclear distance regions over which different vibrational frequency components contribute. This frequency-resolved representation forms the basis of the analysis presented in this work.

The article is organised as follows. The methods section summarizes the initial conditions and approaches used in the AIMS trajectory calculations and electron scattering simulations. The results and discussion section investigates how different vibrational frequencies and the internuclear distance ranges over which they contribute can be extracted from two-dimensional frequency-distance maps. Moreover, we investigate the impact of increasing IRF on the ability to extract meaningful information from these frequency-distance maps. These results are used to examine the impact of methylation on the excited-state dynamics of allene and the corresponding changes observed in the frequency-distance signals. The article concludes with a summary of the main results presented here.

## Methods

### A. Ab initio multiple spawning trajectory simulations

The *ab initio* multiple spawning (AIMS) trajectories and Hessians used in this work were previously reported in Ref. [49], where a detailed description of the calculations is provided. The ground state minimum was optimized, and the corresponding Hessian was computed, at the B3LYP/TZVP level of theory using the Turbomole suite of programs.[78] The excited state trajectories were propagated on surfaces determined on-the-fly at the MR-CIS level of theory using the COLUMBUS package[79]. For both molecules, the underlying complete active space self-consistent field (CASSCF) reference wave functions were constructed using a 4-electron, 4-orbital active space comprising the two highest occupied π orbitals and the two lowest-lying π*orbitals. A 3-state averaging scheme, including the $S_0$, $S_1$(pp*), and $S_2$(pp*) states, was used for both molecules in the CASSCF optimization. The 6-31G** basis set was used for all excited state computations. A total of 39 initial conditions were employed for both the allene and 1,2-butadiene AIMS simulations, drawn from the ground vibrational state Wigner distribution. The equilibrium structures and vibrational frequencies of allene and 1,2-butadiene in their ground-state were additionally calculated with the B3LYP/6-31G* level of theory using QChem.[80]

### B. Electron scattering simulations

Complex atomic elastic electron scattering amplitudes, $f_{\mathrm{i}}(s)$, given by Eqn. (4) for each atomic species were calculated using the ELSEPA package[81] at electron kinetic energies of 90 keV and 3 MeV. For each AIMS geometry, the total elastic electron scattering intensity, $I_{\mathrm{T}}(s,t)$, given by Eqn. (7) was computed within the independent atom model (IAM). At each time step, the total ensemble-averaged electron scattering intensity was obtained by incoherently averaging the electron diffraction signals calculated from the $N_{\mathrm{traj}}(t)$ AIMS geometries at time $t$ using their corresponding normalized amplitude weights, as given by

$$I_T(s,t) = \sum_{\mathrm{k}=1}^{N_{\mathrm{traj}}(t)} I_{T,\mathrm{k}}(s) \cdot w_{\mathrm{k}}(t), \quad (12)$$

where the normalised weight of each molecular geometry, $k$, is given by

$$w_{\mathrm{k}}(t) = |c_{\mathrm{k}}(t)|^2 / \sum_{\mathrm{k}=1}^{N_{\mathrm{traj}}(t)} |c_{\mathrm{k}}(t)|^2, \quad (13)$$

where $c_{\mathrm{k}}(t)$ is the complex amplitude associated with geometry $k$.

In a pump-probe UED experiment, only a fraction, $f_{\mathrm{exc}}$, of the molecules is photoexcited. The diffraction signal from the pumped molecular ensemble is therefore given by

$$I_{\mathrm{pp}}(s,t) = f_{\mathrm{exc}} I_{\mathrm{exc}}(s,t) + (1 - f_{\mathrm{exc}}) I_0(s), \quad (14)$$

where $I_{\mathrm{exc}}(s,t)$ is the scattering signal from the photoexcited ensemble and $I_0(s)$ is the unpumped reference signal calculated from the optimized ground-state equilibrium geometry (see previous sub-section). The corresponding relative difference signal is

$$\Delta I / I_0(s,t) = f_{\mathrm{exc}} [I_{\mathrm{exc}}(s,t) - I_0(s)] / I_0(s). \quad (15)$$

An excitation fraction of $f_{\text{exc}} = 0.1$ was assumed throughout the simulations which is a representative value for gas-phase UED experiments, although the experimentally achieved value depends on the molecular absorption cross-section and pump fluence. A lower excitation fraction reduces the amplitude of the simulated $\Delta\text{PDF}$ and corresponding Fourier-transformed signals but does not change the positions of the retrieved structural or frequency components.

The simulated scattering signal was evaluated over a momentum transfer range of $s = 0 - 15$ Å$^{-1}$ using 256 points to represent the radius of a typical $512 \times 512$ pixel two-dimensional electron detector. A Gaussian convolution of 0.024 Å$^{-1}$ (FWHM) was applied in reciprocal space to approximate the finite momentum transfer resolution arising from the detector pixel size (75 µm) and camera length (0.5 m) for 90 keV electrons. The temporal resolution (i.e., the total IRF) was included by convolving the simulated signals along the pump-probe delay axis, $t$, with a normalised Gaussian of full width at half maximum (FWHM) $\tau$, given by

$$g(t) = (1/\sigma\sqrt{2\pi}) \cdot \exp(-t^2/2\sigma^2), \text{ where } \sigma = \tau/(2\sqrt{2\ln 2}). \qquad (16)$$

Since temporal convolution corresponds to multiplication in the frequency domain, the amplitude of a component oscillating at wavenumber $\nu$ is attenuated by

$$A(\nu, \tau) = \exp(-(\omega\sigma)^2/2), \qquad (17)$$

where $c$ is the speed of light, and $\omega = 2\pi c\nu$.

The difference pair distribution function, $\Delta\text{PDF}(R, t)$, was calculated by sine transformation of $\Delta sM(s, t)$ according to Eqn. (11), using a damping constant of $\alpha = 5 \times 10^{-3}$ Å$^2$. Pair distances were calculated up to $R_{\max}$ = 5 Å. The frequency components of the simulated signals was analysed by Fourier transformation along the pump-probe delay axis, $t$. Before Fourier transformation, the average value of each time-dependent signal over the selected pump-probe delay range was subtracted at each momentum transfer or internuclear distance to remove the zero-frequency (DC) component. A Hann window was subsequently applied to reduce spectral leakage arising from the finite pump-probe delay range, followed by zero padding and Fourier transformed along $t$. The resulting spectral amplitudes, $|\mathcal{F}_t\{\Delta I/I_0\}|(s, \nu)$ and $|\mathcal{F}_t\{\Delta\text{PDF}\}|(R, \nu)$, were evaluated up to $\nu = 3500$ cm$^{-1}$. The latter is referred to throughout this work as the two-dimensional frequency-distance map. We note that the 5-fs time binning of the AIMS trajectories sets a Nyquist limit of 3336 cm$^{-1}$, with frequency components above this limit unable to be reliably retrieved. Moreover, the total pump-probe delay ranges for allene (237 fs) and 1,2-BD (300 fs) limit the frequency resolution to 141 cm$^{-1}$ and 111 cm$^{-1}$, respectively. In addition, restricting the Fourier transform of $\Delta\text{PDF}$ to narrower delay windows of 75-237 fs and 87-300 fs in allene and 1,2-BD, respectively, degrades the frequency resolution to 202 cm$^{-1}$ and 157 cm$^{-1}$.

**Results and Discussion**

## A. Tracking vibrational motions in allene with 5 fs UED

We present the simulated UED signal from the AIMS trajectories of allene initially excited to its $S_1(\pi\pi^*)$ state at 200 nm. Our aim is to investigate how the stretching, twisting, bending, pyramidalization, and $CH_2$ scissoring and wagging vibrational dynamics present in the AIMS trajectories appear in the time-dependent electron scattering signal. Figure 2a shows the corresponding simulated relative difference scattering signal, $\Delta I/I_0$, of allene photoexcited at 200 nm calculated with an IRF of 5 fs.

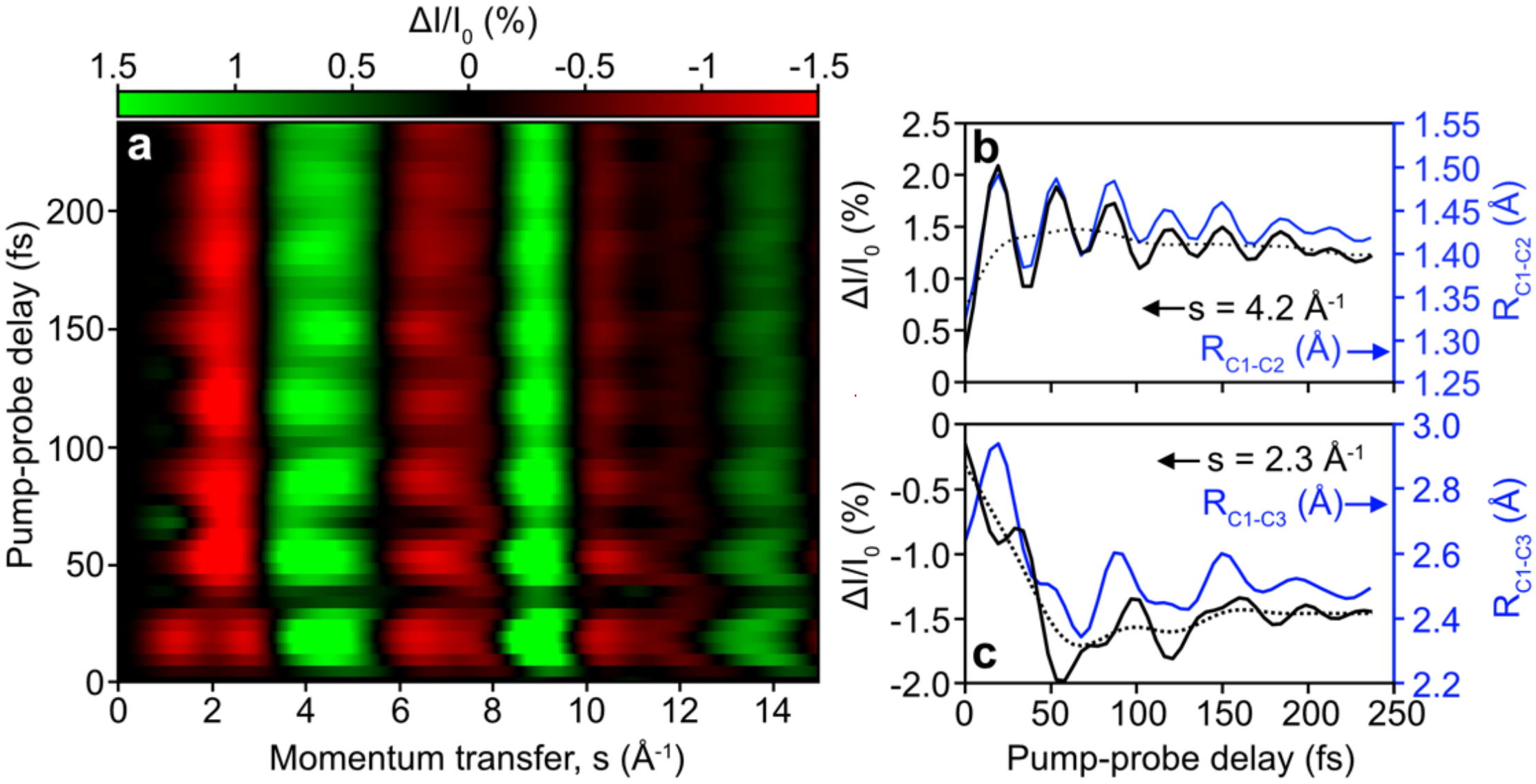

**Fig. 2** **(a)** Two-dimensional distribution of the simulated relative difference electron scattering signal, $\Delta I/I_0$, as a function of momentum transfer, $s$, and pump-probe delay, $t$, of allene following photoexcitation at 200 nm to the $S_1(\pi\pi^*)$ state for an IRF of 5 fs at 90 keV. **(b)** Simulated $\Delta I/I_0$ signal at $s = 4.2$ Å$^{-1}$ (black) with the C1-C2 internuclear distance, $R_{C1-C2}$, from the AIMS trajectories (blue) as a function of pump-probe delay. **(b)** Simulated $\Delta I/I_0$ signal at $s = 2.3$ Å$^{-1}$ (black) with the C1-C3 internuclear distance, $R_{C1-C3}$, which is sensitive to bending of the CCC backbone (blue) as a function of pump-probe delay. The corresponding UED signals calculated with an IRF of 40 fs are also shown by the black dashed lines in panels (b-c).

Oscillatory features are observed at a range of momentum transfers, $s$. For example, the $\Delta I/I_0$ signal at $s = 4.2$ Å$^{-1}$ (black solid, Fig. 2b) has an oscillatory period of 34 fs, with its amplitude decreasing with increasing delay. A similar oscillation is observed in the C1-C2 coordinate, $R_{C1-C2}$, of the AIMS trajectories (blue, Fig. 2b), which oscillates around $R_{C1-C2}$ = 1.43 Å with a period of 34 fs and damps over approximately 200 fs, corresponding to six vibrational periods. The similarity in their oscillation periods suggests that the $\Delta I/I_0$ signal at $s = 4.2$ Å$^{-1}$ may contain a significant contribution from allenic C=C stretching motion.

The C1-C3 internuclear distance, $R_{C1-C3}$, is particularly sensitive to bending of the CCC backbone (blue, Fig. 2c). A significant change in $R_{C1-C3}$ begins after approximately 20 fs, followed by an oscillatory signal containing multiple frequency components. The $\Delta I/I_0$ UED signal at 2.3 Å$^{-1}$ (black, Fig. 2c) shows similar behaviour, with a delayed decrease in the scattering signal followed by more complex oscillatory dynamics. Importantly, increasing the IRF to 40 fs substantially suppresses these oscillatory features (black dashed, Figs. 2b and 2c).

The frequencies of the oscillatory $\Delta I/I_0$ signals were retrieved by Fourier transformation of the $\Delta I/I_0$ signal (Fig. 2a) along its pump-probe delay axis, $t$, with the resulting frequency components as a function of momentum transfer shown in Fig. 3a. Two distinct frequencies are observed at 538 cm$^{-1}$ and 1023cm$^{-1}$. The 538 cm$^{-1}$ component is strongest at lower momentum transfers, predominantly between $s = 1 - 4$ Å$^{-1}$, whereas the 1023 cm$^{-1}$ component extends over a substantially broader momentum transfer range, up to approximately $s = 10$ Å$^{-1}$.

To assign these frequency components to specific nuclear motions, we compare the Fourier spectra of the UED signals with those obtained by Fourier transforming individual internuclear distances of the C1-C2 and C1-C3 coordinates, $R_{C1-C2}$ and $R_{C1-C3}$, respectively, taken directly from the AIMS trajectories. At $s = 4.2$ Å$^{-1}$ (black, Fig. 3b), the UED spectra are dominated by the 1023 cm$^{-1}$ component. This frequency is in good agreement with the corresponding spectra obtained from the time-dependent changes in the C1-C2 internuclear distance, $R_{\mathrm{C1-C2}}$ (blue, Fig. 3b), supporting its assignment predominantly to the allenic C=C stretching motion.

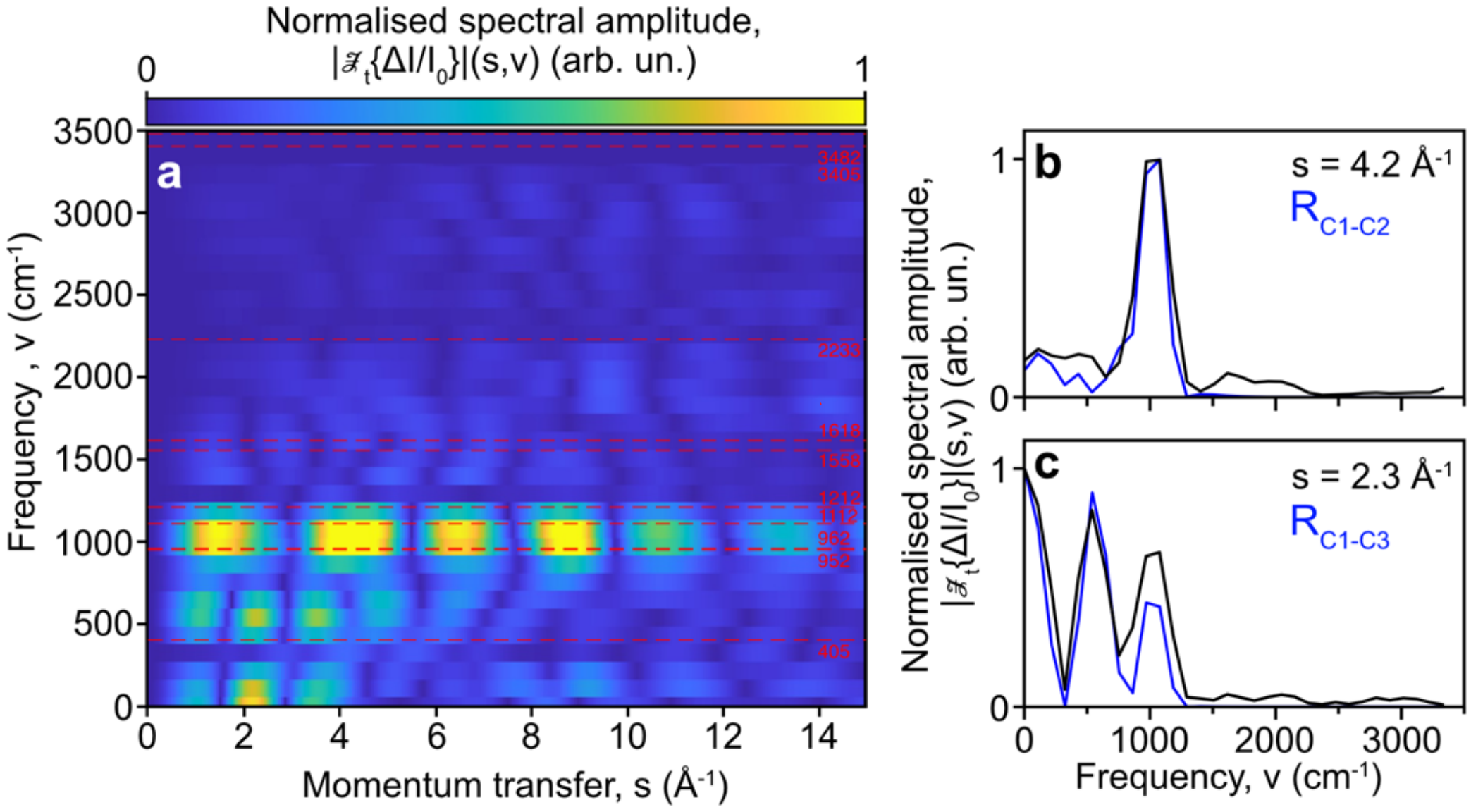

**Fig. 3** **(a)** Two-dimensional distribution of the normalised spectral amplitude, $|\mathcal{F}_{\mathrm{t}}\{\Delta I/I_0\}|(s,\nu)$, as a function of momentum transfer, $s$, and frequency, $\nu$, for allene photoexcited to its $S_1(\pi\pi^*)$ state at 200 nm. The Fourier transformation was performed on the data shown in Fig. 2a along the pump-probe delay axis, $t$, at each momentum transfer. The horizontal red dashed lines indicate the calculated ground-state vibrational frequencies of allene. **(b-c)** Normalised spectral amplitude as a function of frequency, $\nu$, at a selection of momentum transfers (black) of $s = 4.2$ Å$^{-1}$ (b) and $s = 2.3$ Å$^{-1}$ (c), compared with spectra obtained from the time-dependent changes in the C1-C2 (b) and C1-C3 (c) internuclear distances of the AIMS trajectories, $R_{\mathrm{C1-C2}}$ and $R_{\mathrm{C1-C3}}$, respectively.

In comparison, at $s = 2.3$ Å$^{-1}$, the 1023 cm$^{-1}$ component is suppressed, and the spectrum is instead dominated by the 538 cm$^{-1}$ component (black, Fig. 3c). This frequency is in closest agreement with that obtained from the time-dependent C1-C3 internuclear distance, $R_{\mathrm{C1-C3}}$, of the AIMS trajectories (blue, Fig. 3c), which is particularly sensitive to bending of the CCC backbone, supporting the assignment of the 538 cm$^{-1}$ predominantly component to the CCC bending motion. We note that the 1023 cm$^{-1}$ component at $s =$ 2.3 Å$^{-1}$ (black, Fig. 3c) is less strongly suppressed than in the spectrum obtained from $R_{\mathrm{C1-C3}}$ alone (blue, Fig. 3c). This may be due to scattering signal contributions at this momentum transfer from other internuclear distance coordinates. Separating the vibrational motions therefore requires the frequency information to be combined with real-space coordinates, which we address next.

From the $\Delta I/I_0$ scattering signals in reciprocal space and their corresponding frequency analysis, it is challenging to extract further information on the structural dynamics. Therefore, we next investigate the changes in molecular structure through the difference pair distribution function, $\Delta\mathrm{PDF}$. Figure 4a shows the retrieved $\Delta\mathrm{PDF}$ as a function of internuclear distance, $R$, and pump-probe delay, $t$. The nuclear dynamics of excited allene can be described using the $\Delta\mathrm{PDF}$ signals at selected internuclear distances shown in Fig. 4b. In particular, changes in the C1-C3 distance provide information on the CCC bending dynamics, while the C1-C2 and C2-C3 bond distances are sensitive to the C=C stretching motion.

A pronounced diagonal feature is observed in the two-dimensional $\Delta\mathrm{PDF}$ distribution (pink dashed trace, Fig. 4a), which initially follows the stretching of the C=C bonds before evolving towards shorter C1-C3 internuclear distances associated with bending of the CCC backbone. During the first approximately 17 fs, the C=C bonds stretch from 1.3 Å to 1.5 Å following vertical excitation to its $S_1(\pi\pi^*)$ state, with the generated nuclear wavepacket (NWP) reaching the outer turning point at 17 fs. This corresponds to approximately half of the 33-fs period associated with the 1023 cm$^{-1}$ frequency component observed in Fig. 3. The C=C stretching motion is reflected by the depletion of the $\Delta\mathrm{PDF}$ signal around the ground-state C1-C2 and C2-C3 bond lengths of 1.30 Å (red, Fig. 4b) accompanied by a simultaneous rise in $\Delta\mathrm{PDF}$ signal at approximately 1.5 Å (green, Fig. 4b). The signals at 1.3 Å and 1.5 Å subsequently oscillate out of phase with a period of approximately 33 fs.

During the initial 17 fs, the C1-C3 internuclear distance also increases from approximately 2.6 Å to 3.0 Å, giving rise to the positive ΔPDF signal at longer C1-C3 distances (blue and pink, Fig. 4b). After approximately 17 fs, bending of the CCC backbone leads to a contraction of the C1-C3 internuclear distance. This is accompanied by a reduction in the ΔPDF signal at the ground-state C1-C3 internuclear distance of 2.60 Å (blue, Fig. 4b) and a simultaneous increase in signal at 2.23 Å (black, Fig. 4b). The latter is close to the C1-C3 distance of 2.17 Å for the optimized Tw-B $S_1/S_0$ minimum energy conical intersection (MECI) structure. In the two-dimensional ΔPDF distribution, this motion appears as a diagonal feature that initially extends towards $R = 3.0$ Å before contracting towards 2.22 Å and passing through the ground-state C1-C3 distance of 2.60 Å. At delays above 50 fs, the signal around 2.23 Å oscillates with a period of 67 fs, with its amplitude dampening with increasing delay. A C1-C3 distance of 2.23 Å is associated with $\chi \sim 120°$,[49] close to the $S_1$ minimum along the bending coordinate at approximately $130°$ (blue, Fig. 1c).

For the Tw-Py pathway, the pyramidalization dynamics can be probed through changes in the C2-$H_x$ internuclear distances. At the ground-state equilibrium C2-$H_x$ distance of 2.1 Å, a negative ΔPDF signal is observed (orange, Fig. 4b), consistent with the terminal $CH_2$ groups moving away from their ground-state geometry. However, at approximately 70 fs (horizontal grey, Fig. 4b), the signal becomes transiently positive for approximately 20 fs before becoming negative again. This transient coincides with the $S_1$ population decreasing to 50% at 70 fs (blue solid, Fig. 1d) during $S_1 \rightarrow S_0$ internal conversion.

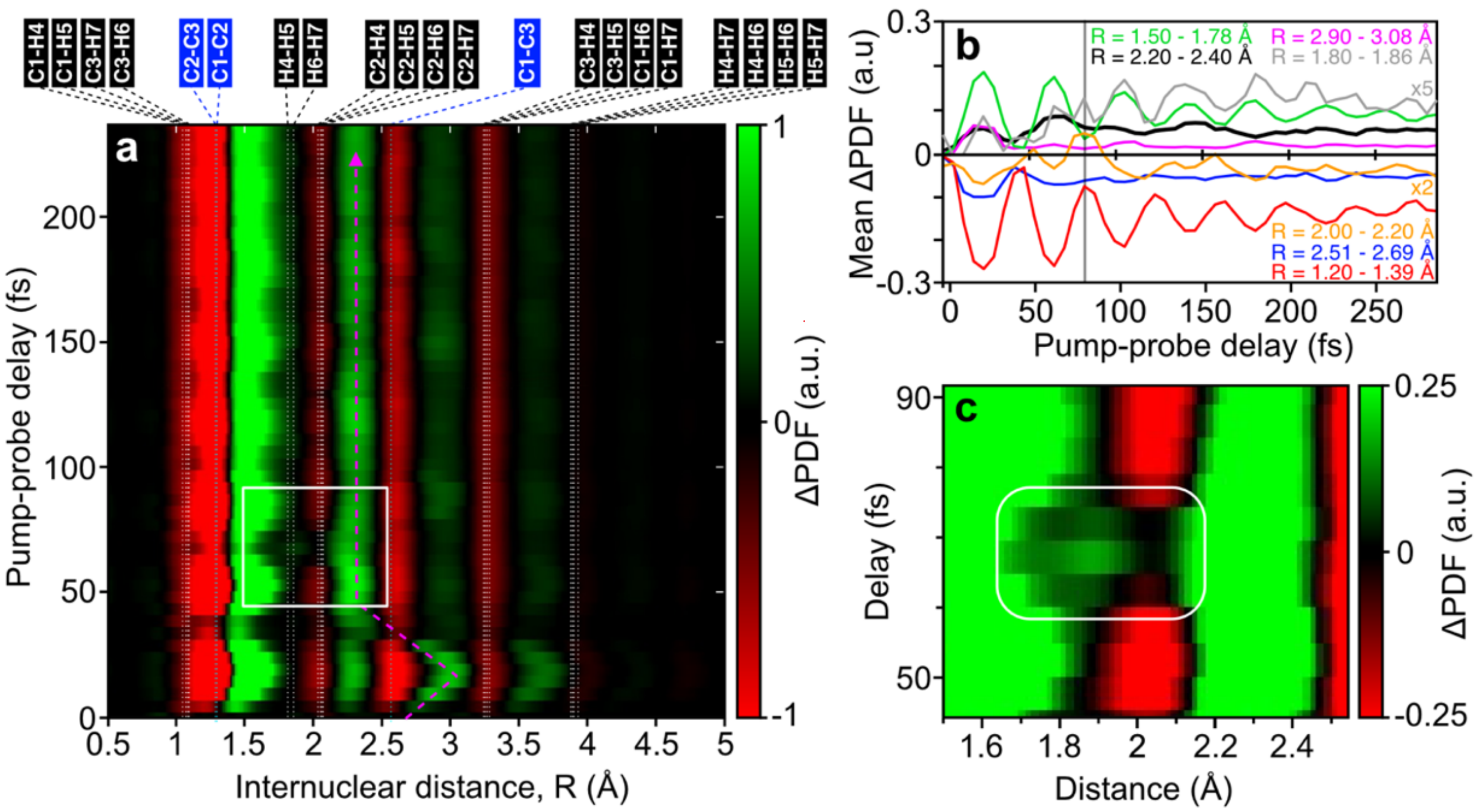

**Fig. 4** **(a)** Two-dimensional simulated difference pair distribution function, ΔPDF, as a function of internuclear distance, $R$, and pump-probe delay, $t$, and of allene following photoexcitation to its $S_1(\pi\pi^*)$ state at 200 nm for an IRF of 5 fs. The vertical dotted lines indicated the ground-state equilibrium internuclear distances, with the corresponding atom pairs labelled. The pink dashed trace highlights the evolution of the C1-C3 internuclear distance associated with the initial C=C stretching and subsequent CCC bending dynamics. **(b)** Mean ΔPDF signals obtained by averaging the ΔPDF signals from panel (a) over selected internuclear distance ranges. The vertical grey lines indicate the appearance of a positive transient at 70 fs for $R = 2.00 - 2.20$ Å (orange). **(c)** Enlarged view of the region highlighted by the white rectangle in (a), showing the transient positive ΔPDF signal between $R = 1.70 - 1.95$ Å at around 70 fs (white rounded rectangle). The colour scale was rescaled to provide greater contrast of the transient feature at approximately 70 fs.

To investigate whether this transient could be associated with sampling of the Tw-Py region of the $S_1/S_0$ intersection seam, we examine the $\Delta\mathrm{PDF}$ signal around the C2-$H_x$ internuclear distances in the optimized Tw-Py $S_1/S_0$ MECI structure. In particular, the C2-H5 distance decreases from its ground-state equilibrium value of 2.07 Å to 1.74 Å at the optimized Tw-Py $S_1/S_0$ MECI structure. The two-dimensional $\Delta\mathrm{PDF}$ distribution shows a positive transient appearing at approximately 70 fs between $R = 1.70 - 1.95$ Å and persisting for approximately 20 fs (see white rectangle in Fig. 4c). The agreement in both internuclear distance and timescale therefore supports assignment of this transient to the NWP sampling the Tw-Py region of the $S_1/S_0$ intersection seam. Importantly, the NWP samples a region of the intersection seam rather than a single optimized MECI geometry, and therefore the range of distances over which the transient appears in Fig. 4c does not need to coincide exactly with the C2-H5 distance in the optimized Tw-Py structure (1.74 Å).

Moreover, with an IRF of 5 fs, it is also possible to resolve the $CH_2$ scissoring motion in allene. The $CH_2$ scissoring dynamics can be probed through the geminal H4-H5 and H6-H7 internuclear distances, which are approximately 1.84 Å in the ground-state equilibrium geometry. Considering the $\Delta\mathrm{PDF}$ signal between 1.80 Å and 1.86 Å (grey, Fig. 4b), the signal remains relatively unchanged during the first 50 fs. After 50 fs, weak high-frequency oscillations appear with a period of 22 fs. The delayed appearance of these oscillations suggests that the contribution from $CH_2$ scissoring becomes more pronounced during the later excited-state dynamics as the NWP evolves away from the Franck-Condon region and towards regions of the $S_1/S_0$ intersection seam.

While the time-dependent $\Delta\mathrm{PDF}$ signal provides direct access to these structural dynamics in real space, identifying the different vibrational frequencies and the internuclear distances over which they contribute requires the individual oscillatory signals to be analysed separately at each internuclear distance. We therefore next combine the frequency and internuclear distance information into a single two-dimensional frequency-distance representation.

### B. Frequency-resolved UED: two-dimensional frequency-distance representation of vibrational dynamics in allene

The two-dimensional frequency-distance representation provides a means of separating the structural dynamics according to both their vibrational frequencies and the internuclear distance regions over which they contribute at. This is particularly useful for several reasons.

Firstly, the overlap of real-space channels observed in the $\Delta\mathrm{PDF}$ distribution makes it challenging to identify nuclear dynamics involving closely spaced internuclear distances. For example, the geminal H4-H5 and H6-H7 distances at 1.83 – 1.86 Å and the C2-$H_x$ distances at 2.04 – 2.08 Å are closely-spaced, making the assignment of the $CH_2$ scissoring and pyramidalization motions difficult from the real-space $\Delta\mathrm{PDF}$ distribution alone. Secondly, a single internuclear distance can be sensitive to more than one vibrational motion. For example, the C1-C3 distance, $R_{\mathrm{C1-C3}}$, depends on both the allenic bond lengths and the CCC bending angle, such that its time dependence contains contributions from both stretching and bending coordinates. Thirdly, distinguishing between symmetric and antisymmetric stretching modes is also challenging from the $\Delta\mathrm{PDF}$ signal alone.

These limitations are addressed by introducing the frequency dimension to the real-space UED signal. We Fourier transform the $\Delta\mathrm{PDF}(R,t)$ distribution along the pump-probe delay axis, $t$, at every internuclear distance, $R$ (Fig. 5a), giving the Fourier components as a function of frequency, $\nu$, and internuclear distance (Fig. 5a). Vibrational motions that overlap in internuclear distance can therefore be separated when they occur at different frequencies, while multiple vibrational motions that contribute to the same internuclear distance can similarly be distinguished according to their frequency components with a 141 $cm^{-1}$ resolution for this 0-237 fs delay range (see Methods).

Two dominant features are present in the frequency-distance map between 860 $cm^{-1}$ and 1200 $cm^{-1}$, centred at the internuclear distances of 1.30 Å and 1.57 Å (white square, Fig. 5a). Although these features occur at two different internuclear distances, they originate from the same C=C bond length coordinate. The feature at 1.30 Å corresponds to depletion of the C=C pair density from its ground-state equilibrium bond length, while the feature at 1.57 Å corresponds to the simultaneous gain in the pair density as the C=C bonds lengthen following photoexcitation (see red and green, Fig. 5b). Both features are centred at a frequency of 1023 cm$^{-1}$, in good agreement with the 33-fs period of the C=C bond-length oscillations observed in the time-dependent $\Delta\mathrm{PDF}$ signals.

The frequency-distance map also provides information on whether this C=C stretching motion is symmetric or antisymmetric. To highlight the depletion and gain contributions in the frequency domain, we use sign-weighted frequency distributions for selected internuclear-distance ranges (Fig. 5c). Here, the frequency spectrum is multiplied by the sign of the corresponding $\Delta\mathrm{PDF}$ signal in Fig. 4a, such that negative and positive values represent depletion and gain contributions, respectively, to the frequency-distance map (Fig. 5a). A depletion at 1.30 Å (red, Fig. 5c) accompanied by a corresponding gain around 1.57 Å (green, Fig. 5c) is consistent with the C1-C2 and C2-C3 bonds lengthening together. In contrast, for an antisymmetric stretch, one C=C bond lengthens as the other shortens, which would broaden or split the frequency-distance distribution around the equilibrium internuclear distance rather than produce the observed concerted depletion at 1.30 Å and simultaneous gain at longer internuclear distance. We therefore assign the 1023 cm$^{-1}$ component predominantly to the symmetric C=C stretching motion. This assignment is also consistent with the symmetry of the initial photoexcited wavepacket. At the $\mathrm{D_{2d}}$ Franck-Condon geometry of allene, the symmetric C=C stretch has $a_1$ symmetry, whereas the antisymmetric stretch has $b_2$ symmetry. Therefore, initial excitation preferentially drives the NWP along the totally symmetric $a_1$ stretching coordinate.

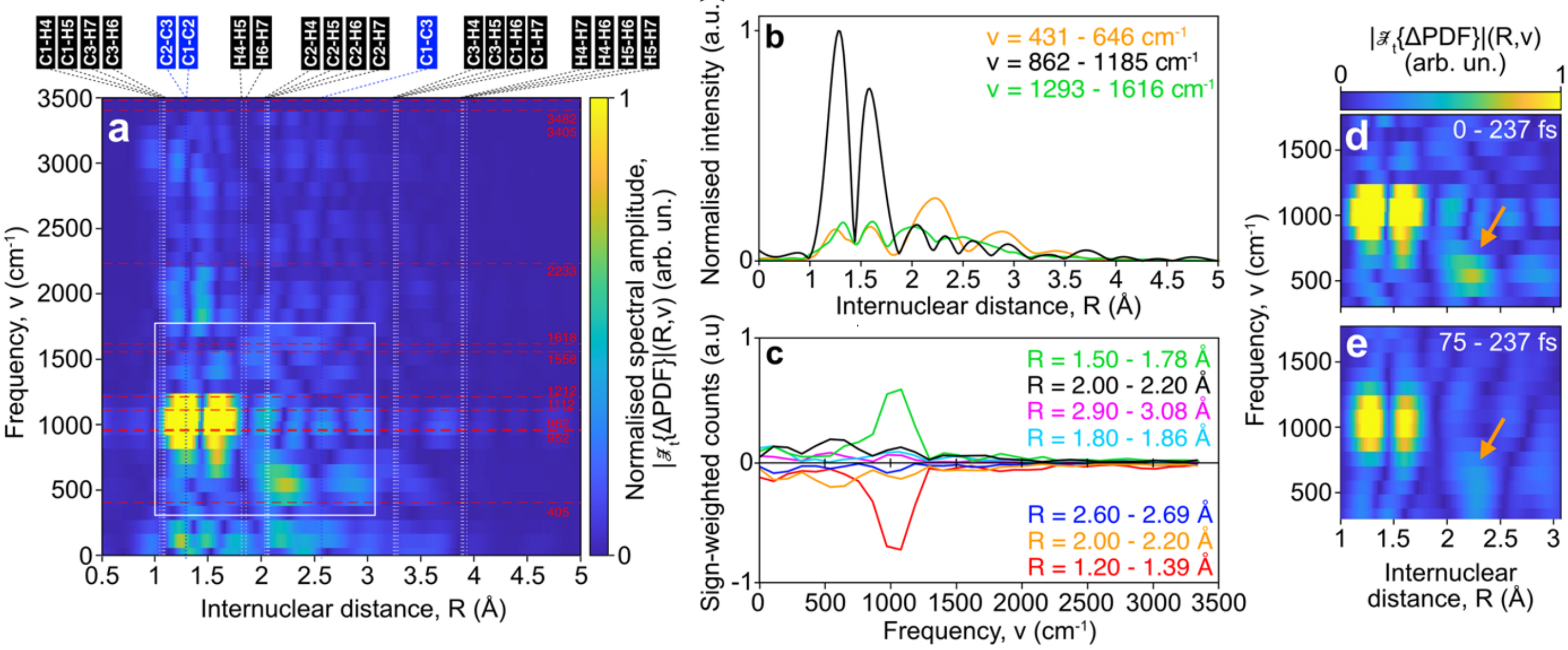

**Fig. 5** **(a)** Two-dimensional distribution of the normalized spectral amplitude, $|\mathcal{F}_t\{\Delta\mathrm{PDF}\}|(R, \nu)$, as a function of internuclear distance, $R$, and frequency, $\nu$, after Fourier transformation of the $\Delta\mathrm{PDF}$ signal in Fig. 4a along the pump-probe delay axis, $t$, for allene excited to its $S_1(\pi\pi^*)$ state at 200 nm. The vertical dotted lines indicated the ground-state equilibrium internuclear distances, with the corresponding atom pairs labelled. The red horizontal lines give the vibrational frequencies of ground-state allene. **(b)** Internuclear distance distribution for selected frequency ranges, $\nu$, extracted from panel (a). **(c)** Sign-weighted frequency distributions for selected ranges of internuclear distance, $R$, generated by multiplying the frequency distributions extracted from panel (a) by the sign of the corresponding $\Delta\mathrm{PDF}$ signal in Fig. 4a. **(d-e)** Enlarged view of the region highlighted by the white rectangle in (a), using pump-probe delay ranges of 0-237 fs (d) and 75-237 fs (e) for Fourier transformation of the $\Delta\mathrm{PDF}$ signal along the pump-probe delay axis. The 538 cm$^{-1}$ bending component is highlighted by the orange arrows.

A second band in the frequency-distance map is centred at 538 cm$^{-1}$, appearing at $R = 2.22$ Å and 2.9 Å (orange square, Fig. 5a). These distances correspond to gain signals in the $\Delta\mathrm{PDF}$ distribution (black and pink, Fig. 4b), accompanied by depletion at the ground-state C1-C3 distance of $R = 2.60$ Å (blue, Fig. 4b). As discussed above, the time-dependent $\Delta\mathrm{PDF}$ distribution shows that the C1-C3 distance initially expands towards 2.9 Å before contracting towards 2.22 Å during the CCC bending dynamics (pink dashed trace, Fig. 4a). The frequency-distance map now shows that this structural evolution contains a strong frequency component centred at 538 $cm^{-1}$, supporting its assignment to the CCC bending motion. The temporal evolution of this frequency component can be further examined by modifying the pump-probe delay range used for the Fourier transformation. When the full 0-237 fs delay is used, the 538 $cm^{-1}$ bending component is clearly observed around $R = 2.2 - 2.4$ Å (orange arrow, Fig. 5d). However, when the early-time dynamics up to approximately the time at which the $S_1$ population decreases to 50% (see grey solid in Fig. 1d) are excluded and the Fourier transform is instead performed over 75-237 fs, this component is strongly depleted (orange arrow, Fig. 5e). The depletion of the 538 $cm^{-1}$ signal after approximately 75 fs coincides with the time at which the $S_1$ population has decreased to 50% (grey solid, Fig. 1d), suggesting that this bending motion is most pronounced during the early $S_1$ dynamics and becomes substantially weaker as the NWP evolves towards and samples regions of the $S_1/S_0$ intersection seam. The 0-75 fs window was not considered further because its short duration results in poor frequency resolution (~450 $cm^{-1}$), limiting meaningful interpretation of the resulting frequency-distance map. Moreover, at the ground-state C1-C3 distance of 2.60 Å (blue, Fig. 5c), a second frequency component at 1023 $cm^{-1}$, associated with the C=C stretching motion, is also present. This demonstrates that a single internuclear distance can contain contributions from multiple vibrational motions (Fig. 5b). Conversely, a single vibrational motion can contribute over multiple internuclear distances as the molecular structure evolves following photoexcitation (Fig. 5c).

Moreover, the frequency dimension allows the closely overlapping $CH_2$ deformation motions associated with $CH_2$ scissoring and wagging to be distinguished, which is challenging from the real-space $\Delta\mathrm{PDF}$ distribution alone. Integrating the frequency-distance distribution over the internuclear distances corresponding to the geminal H4-H5 and H6-H7 pairs, between $R = 1.86 - 1.91$ Å, gives a component at 1508 $cm^{-1}$ and 1618 $cm^{-1}$, and is consistent with the 22-fs oscillation period observed in Fig. 4b. In comparison, the frequency distribution between $R = 1.96 - 2.07$ Å, corresponding to the C2-$H_x$ pairs that are sensitive to terminal-carbon pyramidalization, is dominated by a component at 1077 $cm^{-1}$. This agrees with the calculated $CH_2$ wagging frequency of 1112 $cm^{-1}$, supporting its assignment to wagging motion associated with the pyramidalization coordinate. Such out-of-plane $CH_2$ motion also modifies the C2-$H_x$ distances used above as an observable of terminal-carbon pyramidalization. A weaker contribution from the $CH_2$ scissoring mode is also present over the C2-$H_x$ distance range. The frequency-distance representation therefore enables these overlapping structural contributions to be separated according to their vibrational frequencies, which was not possible from the real-space $\Delta\mathrm{PDF}$ distribution alone.

In addition, several frequencies expected from the calculated ground-state normal modes (red dashed, Fig. 5a) are absent from the frequency-distance map. No significant signal is observed for the C-H stretching modes at 3405 $cm^{-1}$ and 3482 $cm^{-1}$, both of which lie above the Nyquist limit (3336 $cm^{-1}$) and therefore cannot be retrieved. Similarly, no distinct frequency component is observed at the four cross H-H distances (e.g., H4-H7), which lie between $R = 3.89 - 3.93$ Å and are particularly sensitive to twisting of the terminal $CH_2$ groups. We therefore do not identify a distinct frequency-distance signature that can be assigned to the twisting motion, despite the AIMS trajectories showing that twisting plays an important role in carrying the NWP towards regions of the $S_1/S_0$ intersection seam. Nevertheless, as discussed in Section A, the short-lived transient observed around the C2-$H_x$ distances, which are particularly sensitive to terminal-carbon pyramidalization, could provide a real-space signature associated with the NWP sampling of the Tw-Py region of the $S_1/S_0$ intersection seam.

Finally, the frequencies observed during the excited-state dynamics differ from their calculated ground-state normal mode frequencies. The symmetric C=C stretching mode is calculated at 1212 $cm^{-1}$ in the ground state but is observed at approximately 1023 $cm^{-1}$ following photoexcitation, while the CCC bending component is observed at 538 $cm^{-1}$ against a calculated ground-state value of 405 $cm^{-1}$. These two sets of frequencies are not expected to agree. The calculated ground-state values are eigenvalues of the Hessian at the $S_0$ minimum and are therefore harmonic normal mode frequencies defined about a well-defined stationary point. In contrast, the frequencies retrieved from the UED dynamics correspond to frequency components of the nuclear motion as the NWP evolves across the anharmonic excited-state potential, rather than harmonic normal mode frequencies evaluated at a stationary point. These frequencies therefore reflect the curvature and gradients of the regions of the potential energy surface sampled by the NWP and depend on how the wavepacket is prepared. The ground-state frequencies plotted in Figs. 3a and 5a therefore serve to identify which nuclear motions are involved, rather than to predict the frequencies observed during the dynamics. A second consequence is that the observed components need not be equally well defined. A mode that is a spectator to the reaction coordinate continues to oscillate at a reasonably well-defined frequency, whereas a mode that carries the wavepacket towards a CI does not recur over successive periods and is correspondingly broadened. The 538 $cm^{-1}$ bending component belongs to the latter, and is strongly depleted once the first 75 fs are excluded from the Fourier transform of the $\Delta\mathrm{PDF}$ signal, while the 1023 $cm^{-1}$ stretching component persists across the full delay range.

The results presented in this section show how the frequency-distance representation can identify the vibrational motions that contribute to the structural dynamics, particularly the internuclear distances over which they are present. Moreover, we can identify the delay ranges over which specific frequency components are present by selecting different pump-probe delay ranges for the Fourier transformation of the $\Delta\mathrm{PDF}$ signal along the pump-probe delay axis. However, the ability to resolve these frequency components will strongly depend on the temporal resolution of the UED experiment, particularly for the higher-frequency motions. In the next section, we therefore investigate the impact of the total IRF on the frequency-distance maps and the temporal resolution required to resolve the different vibrational motions.

## C. Impact of IRF

So far, the simulated UED results and associated retrievals were presented using the 5-fs time binning of the AIMS trajectories, without any additional temporal convolution. We next investigate the impact of the total IRF on the ability to retrieve meaningful vibrational dynamics in photoexcited allene from UED studies and their extension to frequency-resolved signals. The total IRF was investigated by convolving the simulated signals along the pump-probe delay axis using a Gaussian function of FWHM $\tau$, as given by Eqn. (16). This temporal convolution attenuates the amplitude of a component at a frequency $\nu$, $A(\nu,\tau)$, according to Eqn. (17). The retrieved amplitudes therefore decrease with increasing $\tau$, and the highest-frequency motions being attenuated most strongly.

Figure 6 shows the simulated two-dimensional $\Delta I/I_0$, $\Delta\mathrm{PDF}$, and frequency-distance distributions without additional temporal convolution, corresponding to the 5-fs time sampling of the AIMS trajectories (Fig. 6a), and with a Gaussian temporal convolution with IRFs of 20 fs, 40 fs, and 60 fs (Fig. 6b-d). At an IRF of 20 fs (Fig. 6b), the high-frequency CH2 scissoring component at 1508 cm-1 is no longer clearly visible in the frequency-distance map, with its corresponding oscillatory signature in the $\Delta\mathrm{PDF}$ signal at $R = 1.86 - 1.91$ Å strongly suppressed. The 1023 cm-1 C=C stretching component remains visible at an IRF of 20 fs but is no longer resolved at 40 fs. In comparison, the lower-frequency 538 cm-1 CCC bending component remains weakly visible at an IRF of 40 fs but is lost at 60 fs IRF.

The maximum IRF required to retain at least half of the amplitude of a given frequency component, $\tau_{1/2}$, can be obtained from Eqn. (17) and is given by

$$\tau_{1/2} < (2\ln 2/\pi)\cdot T \approx 0.441\cdot T, \quad (18)$$

where $T$ is the corresponding oscillation period. Table 1 summarizes $\tau_{1/2}$ together with the frequencies and periods of the different vibrational motions. For example, an IRF of approximately 10 fs or shorter is required to retain at least half of the amplitude of the 1508 cm-1 CH2 scissoring component, whereas an IRF of approximately 27 fs is sufficient for the lower-frequency 538 cm-1 CCC bending motion. It is also important to note that transient signal assigned to sampling of the region near and around the S1/S0 intersection seam in Section A persists for only approximately 20 fs and therefore similarly requires a sufficient short temporal response (10 fs or better) to be resolved.

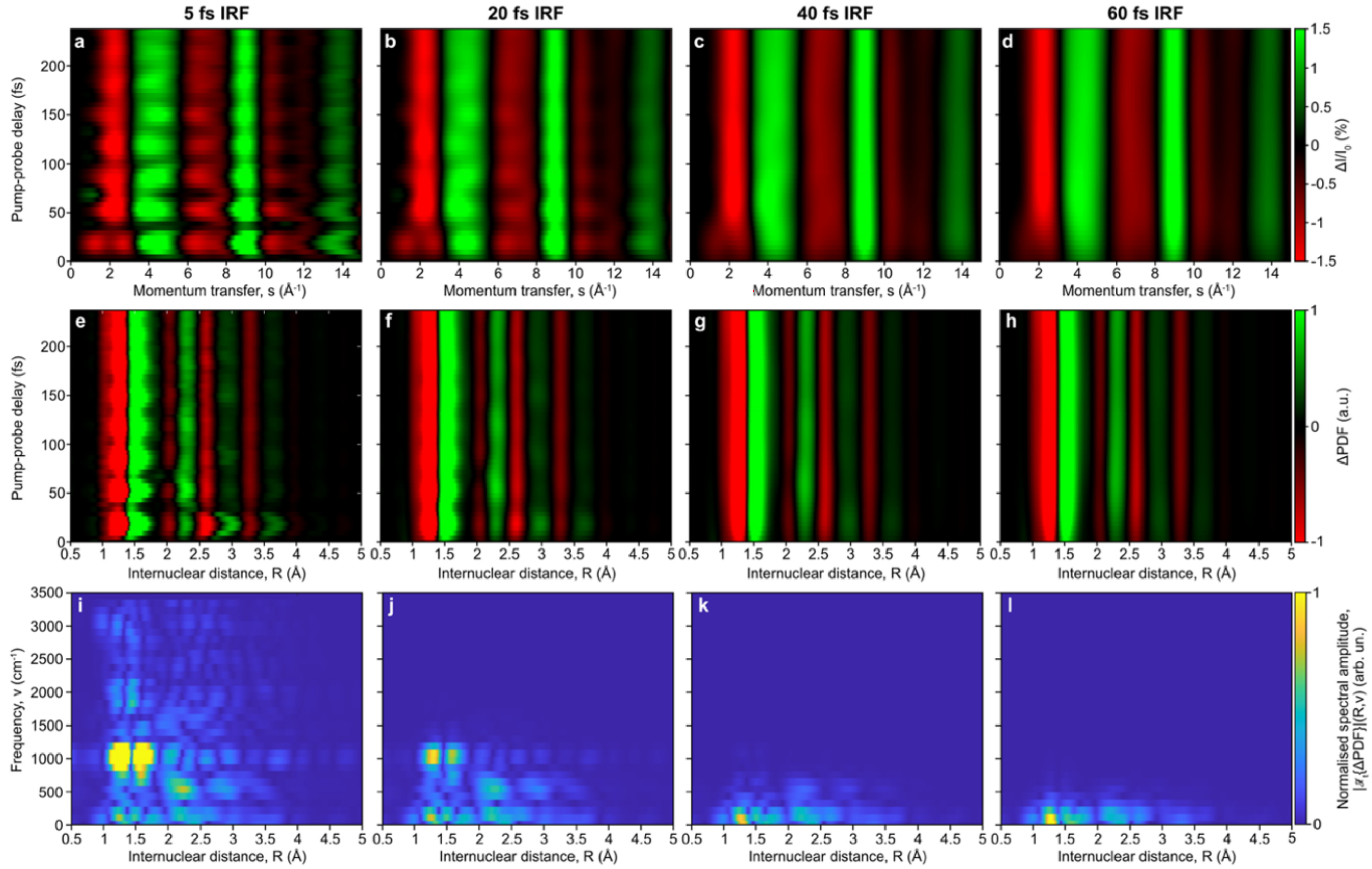

**Fig. 6** Impact of the total instrument response function (IRF; i.e., temporal resolution) on the simulated UED observables of allene following photoexcitation to its $S_1(\pi\pi^*)$ state at 200 nm. **(a-d)** Simulated $\Delta I/I_0$ signal as a function of momentum transfer, $s$, and pump-probe delay, $t$. **(e-h)** $\Delta\mathrm{PDF}$ signal as a function of internuclear distance, $R$, and pump-probe delay, $t$. **(i-l)** Normalized spectral amplitude, $|\mathcal{F}_t\{\Delta\mathrm{PDF}\}|(R,\nu)$, as a function of internuclear distance, $R$, and frequency, $\nu$. The total IRFs are shown for **(a,e,i)** 5 fs, **(b,f,j)** 20 fs, **(c,g,k)** 40 fs, and **(d,h,l)** 60 fs.

While the oscillatory components of the $\Delta\mathrm{PDF}$ signal are progressively lost with increasing IRF, the non-oscillatory structural changes remain observable. At an IRF of 60 fs, the $\Delta\mathrm{PDF}$ distribution still shows depletion and gain signals at several internuclear distances. These features reflect net changes in molecular structure rather than periodic vibrational motion. A measurement with an IRF of 60 fs can therefore still retrieve information on the evolving excited-state molecular structure, while the vibrational dynamics that give rise to these structural changes are strongly suppressed.

**Table 1** Frequencies, corresponding oscillation periods, and IRFs corresponding to 50% attenuation in the oscillation amplitude for the vibrational modes identified in allene following photoexcitation to its $S_1(\pi\pi^*)$ state at 200 nm.

| Mode | Frequency, $\nu$ ($cm^{-1}$) | Period, $T$ (fs) | Minimum IRF, $\tau_{1/2}$ (fs) |
|---|---|---|---|
| CCC bending | 538 | 62.0 | 27.3 |
| C=C stretching, symmetric | 1023 | 32.6 | 14.4 |
| $CH_2$ wagging | 1077 | 31.0 | 13.7 |
| $CH_2$ scissoring | 1508 | 22.1 | 9.8 |
| C=C stretching, antisymmetric (not observed in this work) | 2233* | 14.9 | 6.6 |

*Calculated ground-state normal-mode frequency; no corresponding frequency component was observed in the simulated excited-state UED signal for allene at 200 nm.

Current gas-phase UED instruments typically operate with total IRFs on the order of 80 – 150 fs. At an IRF of 80 fs, even the lowest-frequency component identified here, the 538 $cm^{-1}$ CCC bending motion, is strongly attenuated. The vibrational dynamics identified in Sections A and B would therefore be challenging to resolve directly with the temporal resolution of current gas-phase UED instrumentation. Their observation requires a substantial improvement in the total IRF towards the few-tens-of-femtoseconds and, for the highest-frequency motions, few-femtosecond regime. As discussed in Section B of the Introduction, recent developments in electron-pulse compression and few-fs ultraviolet generation suggest that UED measurements with sub-30-fs, and potentially few-fs, temporal resolution are becoming increasingly feasible.

We next extend this analysis to the methylated derivative of allene, 1,2-butadiene, and investigate its time-dependent electron scattering signals, structural dynamics, and corresponding frequency-distance maps.

### D. Methylated derivative of allene: 1,2-butadiene

Figure 7 presents the simulated UED observables ($\Delta I/I_0$, $\Delta\mathrm{PDF}$, and frequency-distance maps) for the methylated derivative of allene, 1,2-butadiene (1,2-BD), following excitation at 200 nm. Methylation replaces one terminal hydrogen atom with a $CH_3$ group, introducing an additional C-C single bond at 1.51 Å and additional C…C and C…H internuclear distances extending beyond 3.5 Å.

At the ground-state C1-C2 and C2-C3 bond lengths of approximately 1.3 Å, a depletion signal is observed in the $\Delta\mathrm{PDF}$ distribution of 1,2-BD (Fig. 7b; red, Fig. 8a), indicating that these bonds change length following photoexcitation. The frequency-distance map shows two dominant frequency components at 1.3 Å (red square, Fig. 7c) with a 111 $cm^{-1}$ resolution for the 0-300 fs delay range. The dominant frequency at 862 $cm^{-1}$ gives two pronounced features around $R = 1.33$ Å and $1.64$ Å (black, Fig. 8b). The corresponding time-dependent $\Delta\mathrm{PDF}$ signals show depletion around 1.3 Å and gain at 1.6 Å (red and black, Fig. 8a), consistent with substantial elongation of the allenic C=C bonds following photoexcitation.

We assign the 862 $cm^{-1}$ components to the in-phase C=C=C stretching motion, in which C1 and C3 move away from C2 such that both C=C bonds lengthen together. This skeletal C=C=C stretch appears at a lower frequency than in allene (1023 $cm^{-1}$), consistent with methylation of the terminal carbon and the increased mass associated with slower motion of the heavier terminal group in 1,2-BD. The gain signal extending towards 1.64 Å reaches beyond the C1-C2 and C2-C3 bond lengths of the optimized $S_1/S_0$ MECI structures for the Tw-Py (1.33 – 1.36 Å) and Tw-B (1.31 – 1.43 Å) geometries. This suggests that the evolving NWP samples longer C=C bond lengths than those predicted by the static optimized $S_1/S_0$ MECI structures.

A second component centred at 1131 $cm^{-1}$ also contributes within this region of the frequency-distance map. In contrast to the 862 $cm^{-1}$ components, the strongest features of the 1131 $cm^{-1}$ component occur at around $R = 1.28$ Å and $1.49$ Å (green, Fig. 8b). The 1.28 Å distance feature overlaps with the C1-C2 and C2-C3 bond distance region, whereas the feature around 1.49 Å overlaps with the C3-C4 bond introduced by methylation, which has a ground-state equilibrium distance of 1.51 Å. The 1131 $cm^{-1}$ component therefore appears to involve changes over both the allenic C=C and C3-C4 internuclear distance regions. Importantly, this frequency-distance feature is not observed for allene and appears after methyl substitution in 1,2-BD.

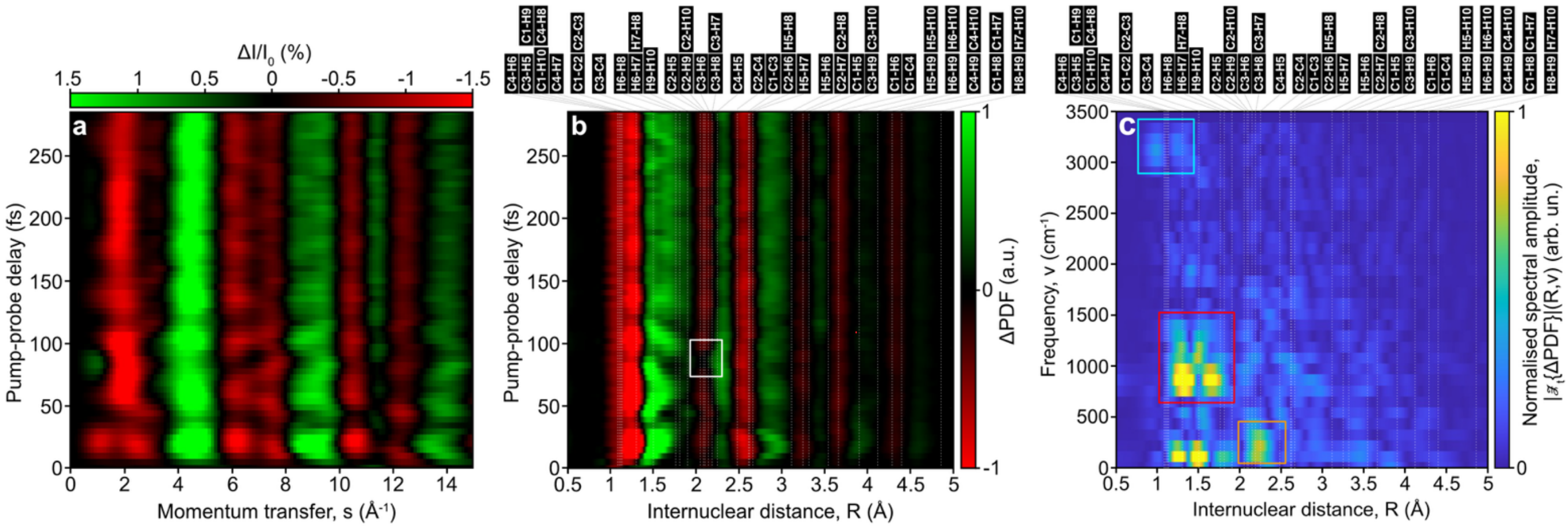

**Fig. 7** Simulated UED signal and structural retrieval from 1,2-butadiene following excitation to its $S_1(\pi\pi^*)$ at 200 nm with an IRF of 5 fs. **(a)** Two-dimensional simulated $\Delta I/I_0$ signal as a function of momentum transfer, $s$, and pump-probe delay, $t$. **(b)** Two-dimensional $\Delta\mathrm{PDF}$ signal as a function of internuclear distance, $R$, and pump-probe delay, $t$. The vertical dotted lines indicated the ground-state equilibrium internuclear distances, with the corresponding atom pairs labelled. **(c)** Two-dimensional distribution of the spectral amplitude, $|\mathcal{F}_t\{\Delta\mathrm{PDF}\}|(R,\nu)$, as a function of internuclear distance, $R$, and frequency, $\nu$. Regions corresponding to the CCC bending (orange square), C=C and C3-C4 internuclear distance changes (red square) and C-H stretching (blue square) are highlighted.

These two higher-frequency components also demonstrate different behaviour depending on the pump-probe delay range used for the Fourier transformation of the $\Delta\mathrm{PDF}$ signal. Restricting the Fourier transform from the full 0-300 fs range to 87-300 fs, with 87 fs chosen to approximately coincide with the time at which substantial $S_1 \rightarrow S_0$ population transfer occurs, retains the dominant 862 $cm^{-1}$ component, whereas the higher-frequency component around 1131 $cm^{-1}$ is strongly depleted (red arrows, Figs. 8d-e). The 1131 $cm^{-1}$ component therefore contributes predominantly during the first 87 fs of the excited-state dynamics, whereas the 862 $cm^{-1}$ skeletal C=C=C stretching component persists to substantially longer pump-probe delays. As discussed above for allene, pyramidalization of the terminal $CH_2$ group towards the Tw-Py region of the $S_1/S_0$ intersection seam produces changes in the C2-$H_x$ internuclear distances, providing a real-space observable of this structural distortion in the $\Delta\mathrm{PDF}$ signal. In 1,2-BD, a short-lived positive transient in around the C2-$H_x$ distance also appears but at a slightly later time (87 fs; orange, Fig. 8a) than in allene (70 fs; orange, Fig. 4b), close to the time at which the calculated $S_1$ population in 1,2-BD decreases to 50% at approximately 90 fs (grey dashed, Fig. 1d). The depletion of the 1131 $cm^{-1}$ component when the first 87 fs are excluded therefore occurs on a similar timescale to substantial nonadiabatic $S_1 \rightarrow S_0$ population transfer. However, the frequency-distance distribution does not show a clear 1131 $cm^{-1}$ contribution at the C2-$H_x$ distances associated with pyramidalization, and we therefore do not directly assign this 1131 $cm^{-1}$ component to the Tw-Py coordinate.

For the bending motion in 1,2-BD, we examine the C1-C3 distance of 2.62 Å, which exhibits a depletion signal in the $\Delta\mathrm{PDF}$ distribution (blue, Fig. 8a). This is accompanied by a gain signal at 2.22 Å (orange, Fig. 8b), which is in good agreement with the C1-C3 distance of 2.25 Å for the optimized Tw-B $S_1/S_0$ MECI structure. A dominant low-frequency component of 215 $cm^{-1}$ is observed around these internuclear distances (orange, Fig. 8a; orange square, Fig. 7c), supporting its assignment to the CCC bending motion. This bending frequency is substantially lower than that observed for allene (538 $cm^{-1}$), consistent with the increased inertia following methyl substitution of the terminal carbon, which slows the bending motion in 1,2-BD, as predicted by Neville et al.[49]

The 215 $cm^{-1}$ bending component also shows a strong dependence on the pump-probe delay range used for Fourier transformation of the $\Delta\mathrm{PDF}$ signal along the pump-probe delay axis. The 215 $cm^{-1}$ component is clearly present when the full 0-300 fs range is considered (orange arrow, Fig. 8d) but is strongly depleted when the first 87 fs are excluded (orange arrow, Fig. 8e). The bending motion therefore contributes predominantly during the early excited-state dynamics. Together with the gain in $\Delta\mathrm{PDF}$ signal around $R =$ 2.22 Å, close to the C1-C3 distance of the optimized Tw-B $S_1/S_0$ MECI structure, this suggests that the early-time bending motion drives the NWP towards the Tw-B region of the $S_1/S_0$ intersection seam.

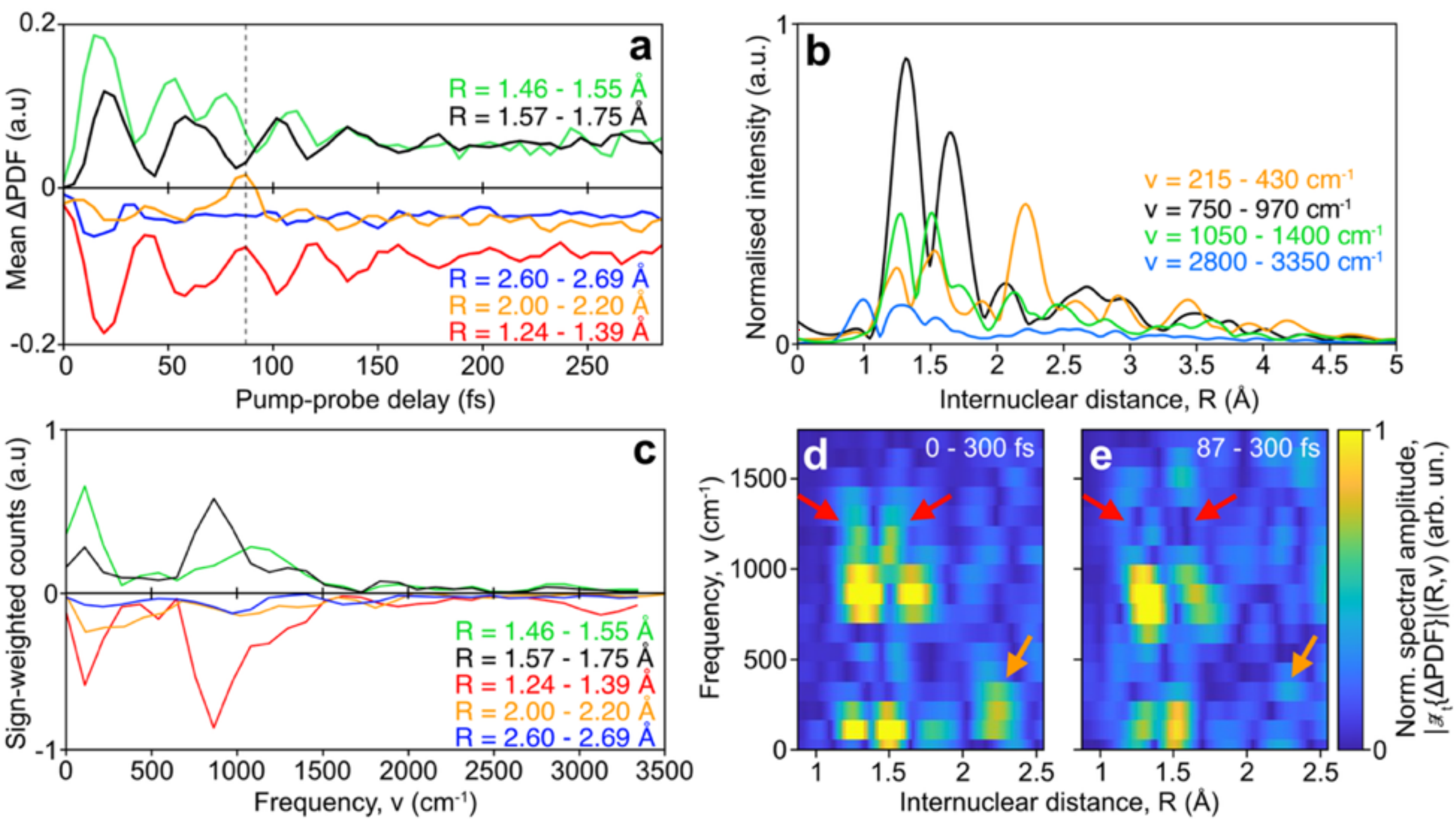

**Fig. 8** **(a)** Mean ΔPDF signal at selected ranges of internuclear distance, $R$, for 1,2-butadiene (1,2-BD) photoexcited to its $S_1(\pi\pi^*)$ state at 200 nm. The time at which the centre of the positive transient signal at $R = 2.00 - 2.20$ Å appears is marked by a grey dashed vertical line. **(b)** Internuclear distance distributions for selected frequency ranges, ν, obtained from the frequency-distance map in Fig. 7c. **(c)** Sign-weighted frequency distribution for selected ranges of internuclear distance, $R$. **(d-e)** Enlarged view of a selected region in the frequency-distance normalised spectral amplitude map, $|\mathcal{F}_t\{\Delta\mathrm{PDF}\}|(R,\nu)$, in Fig. 7c, using the pump-probe delay ranges of 0-300 fs (d) and 87-300 fs (e) for Fourier transformation of the ΔPDF signal along the pump-probe delay axis. The 215 $cm^{-1}$ (orange arrows) and 1131 $cm^{-1}$ (red arrows) frequency components are highlighted, both of which are substantially depleted when the first 87 fs of the dynamics are excluded from the Fourier transformation.

Additionally, a weak high-frequency feature of 3124 $cm^{-1}$ is observed in the frequency-distance map at 1.00 Å and 1.28 Å (blue square, Fig. 7c; blue, Fig. 8b). The feature around 1.0 Å lies close to the C-H bond lengths of 1.09 Å and is therefore assigned predominantly to C-H stretching. The accompanying feature around 1.28 Å lies within the C1-C2 and C2-C3 bond-distance region, indicating that the high-frequency C-H stretching motion also produces a weaker modulation of these pair distances.

The frequency-resolved UED analysis of 1,2-BD shows that methyl substitution modifies both the frequencies and structural coordinates of the vibrational dynamics as compared with allene. For example, the C=C=C stretching frequency decreases in 1,2-BD (862 $cm^{-1}$) compared to allene (1023 $cm^{-1}$), while the CCC bending motion is substantially slowed (215 $cm^{-1}$) relative to its parent molecule (538 $cm^{-1}$). An additional frequency component at 1131 $cm^{-1}$, which is not present in allene, appears following methyl substitution and contributes to the excited-state dynamics predominantly during the first 90 fs of the excited-state dynamics. Similarly, the 215 $cm^{-1}$ bending component also contributes during the first 90 fs, which coincides with substantial $S_1 \rightarrow S_0$ population transfer.

## Conclusions

In conclusion, we introduce frequency-resolved UED, where Fourier transformation of the time-dependent $\Delta\mathrm{PDF}$ signal along the pump-probe delay axis provides a two-dimensional frequency-distance representation of the photoinduced structural dynamics. This frequency-distance representation allows the identification of the vibrational frequencies contributing to the structural dynamics together with the internuclear distances over which they contribute to.

From these maps, we show C=C stretching and CCC bending motions in both allene and 1,2-BD. For the C=C stretching motion in allene, we further demonstrate the ability to distinguish the symmetric C=C stretching contribution. Moreover, methyl substitution leads to the appearance of an additional frequency component that contributes to both the allenic C=C and C3-C4 internuclear distance regions. We find that methylation reduces the frequencies of the C=C stretching and CCC bending motions, consistent with the increased mass of the terminal group and the resulting slower vibrational dynamics. Signals corresponding to $CH_2$ vibrational motions are also identified, including cases where multiple vibrational frequencies contribute at the same internuclear distance.

By modifying the pump-probe delay range used for the Fourier transformation of the $\Delta\mathrm{PDF}$ signal, we further show when specific vibrational frequency components contribute during the photochemical reaction. In both molecules, the CCC bending motion contributes primarily before substantial $S_1 \rightarrow S_0$ population transfer occurs, whereas the C=C stretching component persists over both the earlier and later delay ranges. The new frequency component after methyl substitution also contributes predominantly during the first 90 fs. Moreover, the temporal evolution of some of these frequency components coincides with short-lived (~20 fs) transient structural signals in the $\Delta\mathrm{PDF}$ around the times at which the $S_1$ population decreases to 50% (~70 fs in allene and ~90 fs in 1,2-BD). Finally, we investigate the impact of increasing the total IRF on retaining the vibrational features observed in the frequency-distance maps. Our results show the importance of reaching sub-30-fs total IRFs in UED, and ultimately few-femtosecond temporal resolution, to access the fastest vibrational dynamics in future gas-phase UED experiments. Our results also are not limited to only gas-phase UED but may be extended towards liquid-phase UED as well as potentially to light-matter interactions at interfaces.

## Author contributions

Conceptualization, funding acquisition, project administration, resources, writing – original draft, and supervision: K. A.; data curation, formal analysis, investigation, methodology, validation, visualization, writing – review and editing: R. T., S. P. N., M.S.

## Conflicts of interest

There are no conflicts to declare.

## Data availability

The data supporting the findings of this study are available from the corresponding author upon reasonable request. The AIMS trajectory data are not currently deposited in a public repository because they form part of ongoing and unpublished studies by the authors. These data will be made publicly available following completion and publication of the associated work.

## Acknowledgements

K. A. acknowledges financial support from the European Research Council through ERC Starting Grant “TERES” (Grant No. 101165245), Lasers4EU (Grant No. 101131771, Project ID 37011), and the Deutsche Forschungsgemeinschaft (DFG, German Research Foundation) through Sonderforschungsbereiche (SFB) 1477 “Light-Matter Interactions at Interfaces (LiMatI)” (Project No. 441234705). R. T. acknowledges support from Paris-Saclay Institut d'Optique Graduate School. M. S. acknowledges the support of the NSERC Discovery grant program.

Funded by the European Union. Views and opinions expressed are however those of the author(s) only and do not necessarily reflect those of the European Union or the European Research Council Executive Agency. Neither the European Union nor the granting authority can be held responsible for them. This work is supported by ERC grant (TERES, 101165245, Project DOI: 10.3030/101165245).